\documentclass[pra,aps,a4paper, twocolumn,repreprint, superscriptaddress, longbibliography]{revtex4-2}

\usepackage{amssymb}
\usepackage{amsmath}
\usepackage{amsfonts}
\usepackage{graphicx}
\usepackage{color, soul}
\usepackage{natbib}
\usepackage{hyperref}
\usepackage{adjustbox}
\usepackage{tabularx}
\usepackage{breqn}
\usepackage{multirow}

\usepackage{lineno}

\begin{document}

     \title{Phonon heat capacity and disorder: new opportunities for performance enhancement of superconducting devices}

	\author{M. Sidorova}
    \email{Mariia.Sidorova@dlr.de}
    \affiliation{Nanyang Technological University$,$School of Physical and Mathematical Science $,$ 21 Nanyang Link $,$ 637371 Singapore}
	\affiliation{Humboldt-Universität zu Berlin$,$ Department of Physics$,$ Newtonstr. 15$,$ 12489 Berlin$,$ Germany}
	\affiliation{German Aerospace Center (DLR)$,$ Institute of Optical Sensor Systems$,$ Rutherfordstr. 2$,$ 12489 Berlin$,$ Germany}
	
	\author{A.D. Semenov}
	\affiliation{German Aerospace Center (DLR)$,$ Institute of Optical Sensor Systems$,$ Rutherfordstr. 2$,$ 12489 Berlin$,$ Germany}
	
	\author{I. Charaev}
	\affiliation{Physics Institute$,$ University of Zürich$,$ Winterthurerstrasse 190$,$ 8057 Zürich$,$ Switzerland}

    \author{M.  Gonzalez}
	\affiliation{Physics Institute$,$ University of Zürich$,$ Winterthurerstrasse 190$,$ 8057 Zürich$,$ Switzerland}

	\author{A. Schilling}
	\affiliation{Physics Institute$,$ University of Zürich$,$ Winterthurerstrasse 190$,$ 8057 Zürich$,$ Switzerland}

    \author{S. Gyger}
	\affiliation{Department of Applied Physics$,$ KTH Royal Institute of Technology$,$ SE-106 91 Stockholm$,$ Sweden}
    \affiliation{Ginzton Laboratory$,$ Stanford University$,$ 348 Via Pueblo Mall$,$ Stanford$,$ California 94305$,$ USA}
	
	\author{S. Steinhauer}
	\affiliation{Department of Applied Physics$,$ KTH Royal Institute of Technology$,$ SE-106 91 Stockholm$,$ Sweden}

	\begin{abstract}
	
	We experimentally study for the first time the self-heating normal domain in magnetic field using nanowires made of granular NbTiN films with sub-10~nm thicknesses. Specifically, we found that at temperatures below 10 K, the heat capacity of phonons in such films is reduced with respect to predictions of the Debye model. Moreover, as the temperature decreases, the phonon heat capacity reduces quicker than the Debye prediction. These effects strengthen when the film thickness decreases. We attribute the observed reduction in the heat capacity to the size effect, which arises from the confinement of phonon modes within the grains; a phenomenon that has been predicted but not yet observed. These findings hold great importance in understanding the role of heat transport in superconducting electronic devices and have the potential for practical applications in mid-infrared photon sensing with superconducting nanowire detectors.
	
	\end{abstract} 

\date{\today}
\maketitle

\section{Introduction}
\label{sec: Introduction}

Controlling phonon heat capacity and disorder in superconducting materials forms a scientific-technological basis for the performance enhancement of  various superconducting electronic devices, particularly thermal switches (nanocryotrons) \cite{mccaughan2014superconducting, baghdadi2020multilayered} and single-photon detectors \cite{dello2022advances}. For instance, in Superconducting Nanowire Single-Photon Detectors (SNSPDs) employed, e.g., in space communications \cite{you2018superconducting}, the sensitivity to low-energy photons, i.e. the quantum yield, $\eta$, is controlled by various parameters including phonon heat capacity. More specifically, the sensitivity grows exponentially with the fraction of the photon energy transferred to the electron system, $\eta = c_e/(c_e+c_{ph})$ \cite{vodolazov2017single}, where $c_e$ and $c_{ph}$ are the electrons and the phonon heat capacities, respectively. Thus, reducing $c_{ph}$ can serve as a means of enhancing SNSPDs' sensitivity towards far-infrared photons. Another example are Hot Electron Bolometers (HEBs) employed, e.g., in terahertz spectroscopy, which basically detect a temperature change due to the heating of electrons by absorbed radiation. The responsivity of HEB-mixers - a key metric in their numeric calibration \cite{Baselmans2006DirDet} - is affected by various material parameters including the $c_e/c_{ph}$ ratio. Any inaccuracies in defining the $c_e/c_{ph}$ ratio eventually lead to calibration errors, thereby altering the retrieved width and amplitude of gas lines measured using HEBs. For example, in the case of atomic oxygen \cite{richter20154}, which plays an important role in the chemistry and energy balance of the earth's atmosphere, any such errors can skew the forecast for climate change.

Another intriguing effect of phonon heat capacity, which can be used to mitigate noise in various superconducting devices, is related to thermal fluctuations in electron energy. Given that phonons mediate the electron-to-thermal bath coupling, a reduction in $c_{ph}$ leads to a decrease in the variance of electron energy fluctuations, in full accordance with statistical thermodynamics \cite{hajdu1977r}. This might play a key role in reducing the dark count rate of SNSPDs \cite{semenov2020local}, a critical metric when detecting extremely low and rare events such as arrivals of dark matter particles \cite{chiles2022new}.

A way to control $c_{ph}$ is the reduction of the device dimensionality in one, two, or all three directions as in films, wires, or grains that produces two important so-called size effects \cite{halperin1986quantum}. As the dimensionality is reduced, the number of allowed phonon wavevectors (density of phonon states) is also reduced due to confinement of phonon modes. Simultaneously, the surface-to-volume ratio increases that leads to an emergence of surface modes contributing to the heat capacity. The theory \cite{baltes1973specific, lautehschlager1975improved, nishiguchi1981vibrational, prasher1999non, mcnamara2010quantum} predicts an enhancement of $c_{ph}$ in nanostructures with free (suspended) surfaces or, otherwise, its reduction when surfaces are clamped, i.e., when the surface modes are prohibited. To the best of our knowledge, only enhancement of the phonon heat capacity has been reported. Furthermore, as the device dimensionality is reduced, other effects come into play, such as an increase in the level of disorder. The increase in the disorder affects other film properties and finally the device performance. For example, disorder-driven reduced diffusivity and superconducting gap are beneficial for the low-energy photon sensitivity of SNSPDs. To some extent, the phonon heat capacity anti-correlates with the level of disorder which makes the task of manipulating only one parameter non-trivial.

Contemporary nanotechnology offers cutting-edge techniques to reduce device dimensions down to the atomic scale and control the disorder. Methods such as molecular-beam epitaxy (MBE) \cite{wright2021unexplored} and atomic-layer deposition (ALD) \cite{klug2011atomic, shibalov2021multistep} allow controlling the thickness of sputtered material down to a single atomic layer. Additionally, electron-beam lithography on ultra-high resolution resists \cite{grigorescu2009resists} with sub-nm helium ion beam \cite{he2020helium} breaks the sub-10~nm restriction \cite{rommel2013sub} in the fabrication of nanostructures. Even in a post-fabrication process, material properties can be tuned in a controllable and reproducible manner, e.g., by means of dislocation engineering, the superconducting transition temperature and diffusivity can be changed to predefined values \cite{livengood2009subsurface, Kim_2020, cybart2015nano, zhang2019saturating}. Moreover the grain size in granular films can be adjusted by manipulating the sputtering conditions such as ambient temperature and gas flow \cite{dane2017bias}.   

In this study, we employ a well-established technique of creating self-heating normal domain \cite{skocpol1974self, yamasaki1979self} in order to investigate the behavior of phonon heat capacity in sub-10~nm NbTiN films \cite{sidorova2022phonon}. We show that the heat capacity is disorder-dependent and is reduced with respect to predictions of the Debye model.

\section{Experimental approach}
\label{sec: Section name}

\begin{figure}[t!]
	\centerline{\includegraphics[width=0.45\textwidth]{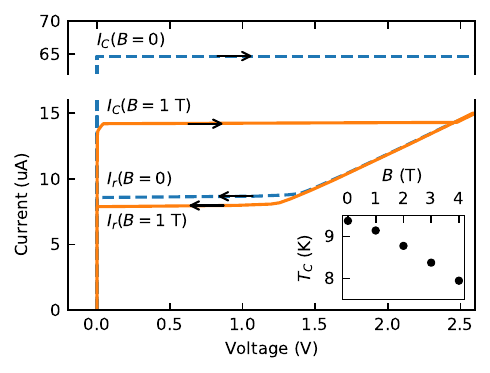}}
	\caption{IV curves  measured at the bath temperature $T_B=2.4$~K with ($B = 1$~T, solid line) and without ($B = 0$, dashed line) external magnetic field for the wire with the thickness $d=9$~nm. Arrows indicate the sweep direction in the current-bias regime. $I_C$ and $I_r$ are switching and hysteresis (return or retrapping) currents discussed in the text. Inset: Field-dependent transition temperature.}
	\label{fig:IV}
\end{figure}

Two superconducting NbTiN films with thicknesses $d = 6$~nm and 9~nm were fabricated on Si substrates above 270~nm-thick thermally grown SiO$_2$ buffer layers (properties of exactly these films were thoroughly studied in \cite{sidorova2021magnetoconductance}). The films were patterned into straight nanowires with widths $w=200$~nm and lengths $l=100~\mu$m and $80~\mu$m, respectively. Measurements were conducted in a physical-property measurement system (PPMS, Quantum Design) in different magnetic fields ranging from 0~T to 4~T. The field was applied perpendicular to the film surface.

Fig.~\ref{fig:IV} shows typical hysteretic current-voltage ($IV$) curves of a superconducting nanowire in zero magnetic field $B=0$ (dashed curve) and in the field $B = 1$~T (solid curve). In the current-bias mode, sweeping the current from zero upwards switches the wire from the superconducting to the normal state at the experimental critical (switching) current $I_C$. Further sweeping the current downwards returns the wire back to the superconducting state through a resistive plateau defined by the hysteresis (return or retrapping) current $I_r$. The plateau is linked with a self-heating normal domain with a length proportional to the voltage. Magnetic field reduces both currents, $I_C$ and $I_r$ (solid curve in fig.~\ref{fig:IV}). It also suppresses the superconducting transition temperature, $T_C$, defined here as the temperature at which the wire resistance equals one half of its normal value at about 25~K (inset in fig.~\ref{fig:IV}). We note here that the measured $I_r(B=0)$ for the 9~nm-thick wire is consistent with values reported earlier for similar 10~nm-thick NbTiN films \cite{steinhauer2020nbtin} (see also \cite{Ir_substrates}).

\begin{figure}[t!]
	\centerline{\includegraphics[width=0.48\textwidth]{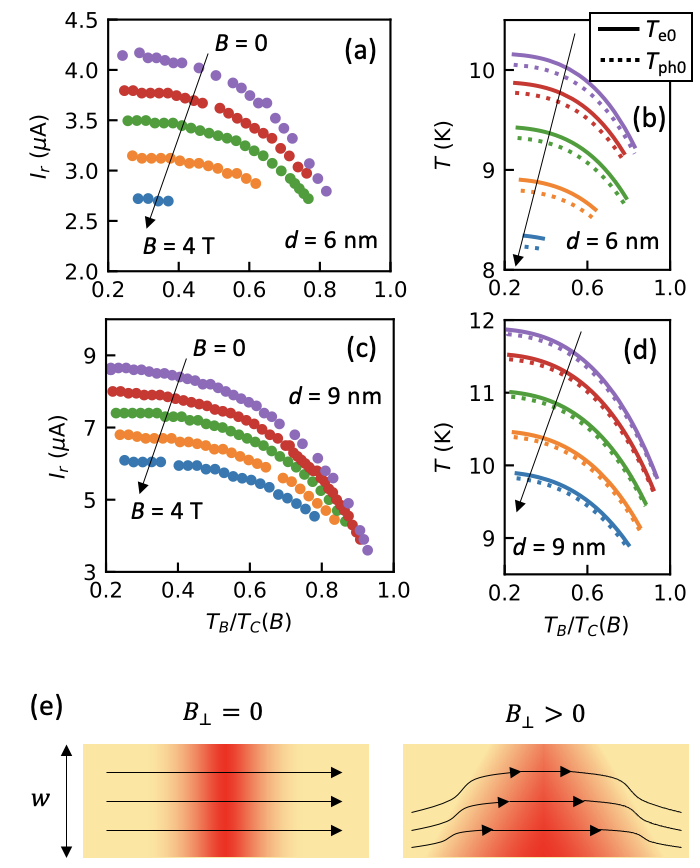}}
	\caption{Experimental hysteresis currents at preset magnetic fields ($B$ = 0, 1, 2, 3, and 4~T) vs. bath temperatures ($T_B$) normalized with field-dependent transition temperatures $T_C(B)$ for 200~nm-wide wires  with thicknesses (a) $d=9$~nm, and (c) 6~nm. (b) and (d) Show electron $T_{e0}$ and phonon $T_{ph0}$ temperatures in the normal domain center for these wires. 
 Arrows indicate an increase in the magnetic field. (e) Schematics of the effect of an external magnetic field on the current density (presented by current lines) and on the domain shape (shown with red color). For large domains, variations in the domain edges do not affect $T_{e0}$ and $T_{ph0}$.}
	\label{fig:Ir}
\end{figure}
In fig~\ref{fig:Ir}, we plot $I_r(T_B, B)$ dependences which were extracted from $IV$ curves measured at different bath temperatures, $T_B$, and preset magnetic fields. At $B = 0$, the $I_r (T_B, B=0)$ dependence has been theoretically described \cite{sidorova2022phonon} with the model of the self-heating normal domain. The hysteresis current is given by
\begin{equation}\label{eq:current}
I_r = \sqrt{\frac{c_e(T_{e0}) w^2 d }{p\:\tau_\epsilon(T_{e0}) R_\square  T_{e0}^{p - 1}}(T_{e0}^p - T_B^p)},
\end{equation}
where $T_{e0}$, $c_e$, $\tau_\epsilon$, and $R_\square$ are the electron temperature in the domain center, the volumetric heat capacity of electrons, the relaxation time of the electron energy to the substrate, and the sheet resistance of the film, respectively.  The temperature in the domain center $T_{e0}$ for each pair of values of $B$ and $T_B$ is determined as numerical solution of the algebraic equation
\begin{eqnarray}\label{eq:t_J_t_B} 
	T_B^{p+1} + \dfrac{p+1}{p} (T_{e0}^p-T_B^p) T_C(B) - T_{e0}^{p+1} = 0
\end{eqnarray}
which follows from the
heat balance equations for electrons and appropriate boundary conditions 
(details are given in \cite{sidorova2022phonon} and briefly highlighted in appendix~\ref{app_SHNM}). Here $T_C(B)$ is the transition temperature in the applied magnetic field. The values of $T_{e0}$ obtained from eq.~\ref{eq:t_J_t_B} are shown in  fig~\ref{fig:Ir} (b) and (d). There are also shown effective phonon temperatures $T_{ph0}$ in the domain center. They were additionally computed in the framework of the uniform, non-linear two-temperature (2T) model \cite{comment_2TM}. Although the difference between these two temperatures, $T_{e0}-T_{ph0}$, is small compared to $T_{e0}$ and can be neglected for practical purposes, we accounted for it in the following evaluation of experimental data.

The fields used in our experiment (1.0~T - 4.0~T) drive the superconducting wire into a mixed vortex state, emerging between the upper-critical field $B_{c2}(T=0)>13.9$~T (measured for the studied NbTiN films in \cite{sidorova2021magnetoconductance}) and the field $B_C\approx0.2$~T at which a transition from the vortex-free Meissner to mixed vortex state occurs (estimated according to the Ginzburg–Landau model \cite{maksimova2001critical}). The estimated $B_C$ is in consistence with previously reported values for similar NbTiN films \cite{jonsson2022current}. Furthermore, a non-zero magnetic field induces screening currents which sum up with the transport current. This makes the net current density across the superconducting part of the wire non-uniform as illustrated in fig.~\ref{fig:Ir}(e) with current lines. Although the non-uniformity is expected to alter the shape of domain edges, for large domains the translational symmetry (space homogeneity) with respect to domain edges dictates that the edges and the mixed state of superconducting parts of the wire do not affect the temperature at the domain  center $T_{e0}$. Moreover, for sufficiently large domains, finite temperature at the domain center requires the derivative $\partial^2 T_e/\partial x^2$ at the center to approaches zero. These conditions allow one to apply for a non-zero magnetic field the formalism which we developed earlier \cite{sidorova2022phonon} for $B = 0$ in the framework of the model of a self-heating normal domain.


To validate the use of eq.~(\ref{eq:current}) for the evaluation of the experimental data at $I_r(T_B, B>0)$, we first estimate the impact of magnetic field on the parameters entering this equation. Magnetic field separates the electron states with opposite spin directions by the Zeeman splitting energy $2 \mu_B B$ ($\mu_B$ is the Bohr magneton) that affects the electron density of states $D(E)$ and, consequently, the electron heat capacity $c_e=\int D(E)f_{FD}(E)E dE$ (here $f_{FD}(E)$ is the Fermi-Dirac distribution function). The correction to the electron heat capacity is of the order of $(\mu_B B/E_F)^2/8$ 
where $E_F$ is the Fermi energy.  For fields used in our experiment, the splitting energy ($<0.5$~meV) is much smaller than the Fermi energy ($\approx 5$~eV), and therefore the effect of splitting on $D(E)$ and $c_e$ is negligible. 

We associated $\tau_\epsilon(T_{e0})$ introduced in eq.~(\ref{eq:current}) with the photoresponse time defined with the modified 2T model \cite{comment_2TM} similar to the conventional, linear 2T model in \cite{perrin1983response}. In both models, the response time is a function of the ratio $c_e(T_e)/c_{ph}(T_{ph})$, the phonon escape time $\tau_{esc}$, and the electron-phonon (\textit{e-ph}) energy relaxation time $\tau_{EP}(T_e)$. Obviously, magnetic field has no direct impact on the number of phonons and, therefore, no effect on $c_{ph}$ and $\tau_{esc}$ and, as it is shown above, has a negligible effect on $c_e$. In order to evaluate the effect of magnetic field on $\tau_\epsilon(T_{e0})$ one has to consider possible field effects in $\tau_{EP}$.

In a non-magnetic and non-piezoelectric normal metal, magnetic field can affect the \textit{e-ph} coupling via screening of electromagnetic interaction between electrons and vibrating ions of the crystalline lattice and via induced changes in the deformation potential. The effect through both channels is parameterized with the frequency-dependent conductivity tensor \cite{SPECTOR1967291}. The direct electromagnetic interaction is completely screened in our case, since, for $T_{ph0} = 10$~K, frequencies of thermal acoustic phonons $\omega_T/(2\pi) \approx 2\cdot10^{11}$~Hz are much less than the plasma frequency $\omega_p/(2\pi) \approx 4\cdot10^{15}$~Hz and the characteristic frequency $\sigma^{-1}\varepsilon_0  \varepsilon_r/(2\pi)\approx 2\cdot 10^{15}$~Hz ($\sigma$ and $\varepsilon_r$ are the dc conductivity and  the dielectric permittivity at infinite frequency, respectively) above which the conductivity becomes frequency dependent. 
On the other hand, the applied magnetic fields are not strong enough to cause cyclotron resonance (Landau quantization). Indeed, in our case the cyclotron frequency $\omega_c = e B/m_e$ ($\omega_c/(2\pi) \approx 1.1\cdot10^{11}$~Hz for $B=4$~T) is much smaller than the reciprocal elastic scattering time $\tau\approx 1$~fs \cite{sidorova2021magnetoconductance}. 
Under these conditions, electrons do not complete at least one entire orbit without being scattered and, consequently,
cannot experience a resonance-energy exchange with thermal phonons. Corrections to the diagonal elements in the conductivity tensor \cite{abrikosov2017fundamentals} are of the order $(\omega_c\tau)^2 \approx 1.5\cdot10^{-6}$ and can be safely neglected.
Finally, since all the discussed material parameters do not depend on the magnetic field, the exponent $p$ in eq.~(\ref{eq:current}) is likewise field-independent (in a particular film $p$ is determined by the phonon dimensionality and the degree of disorder \cite{bezuglyi1997kinetics}).

\begin{figure}[t!]
	\centerline{\includegraphics[width=0.45\textwidth]{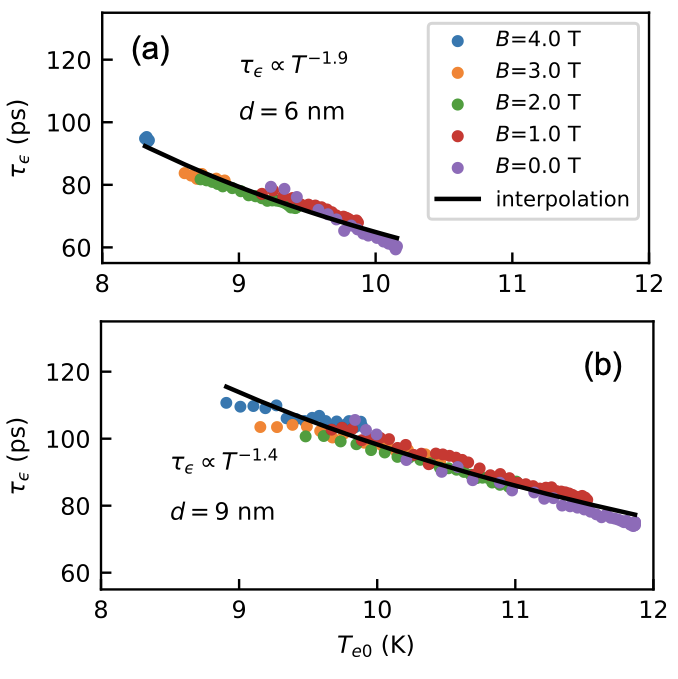}}
	\caption{Electron energy relaxation time vs. temperature in the domain center for the wires with (a) $d=6$~nm and (b) $d=9$~nm, respectively. Symbols: the best-fit values obtained by fitting experimental $I_r(T_B)$ data with eq.~(\ref{eq:current}), colors indicate magnetic fields listed in the shared legend; black solid curves: the best fits to these data with a power-law function indicated in the labels.}
	\label{fig:taus}
\end{figure}

\begin{figure}[t!]
	\centerline{\includegraphics[width=0.48\textwidth]{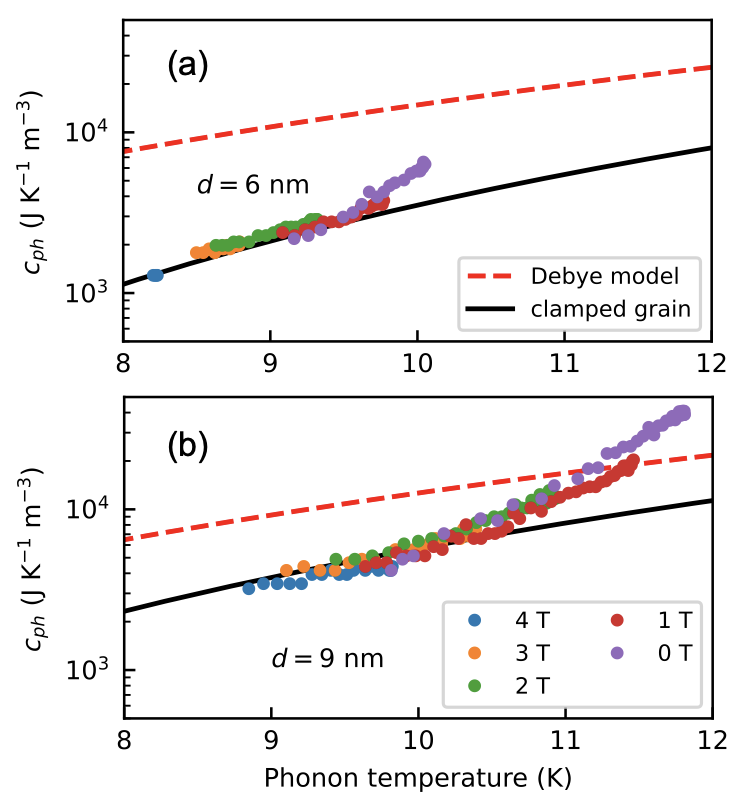}}
	\caption{Heat capacity of phonons vs. phonon temperature in the domain center for two NbTiN wires with thicknesses (a) 6~nm and (b) 9~nm; figures share legends. Scatters: experimental values extracted from $\tau_\epsilon$ (colors indicate different magnetic fields); dashed curves: predictions of the Debye model for bulk material; solid curves: best fits for a cubic grain with the clamped surface obtained with eq.~(\ref{eq:c_particle}) with a size of 1.7~nm and 3~nm for the thin and thick wires, respectively.}
	\label{fig:cphs}
\end{figure}

Assuming field-independent material parameters, we fit the experimental $I_r(T_B, B)$ data shown in fig.~\ref{fig:Ir}(a) and (c) with eq.~(\ref{eq:current}). The temperatures $T_{e0}$ were found from eq.~(\ref{eq:t_J_t_B}) for each pair of values $B$ and $T_B$. The values of $c_e$ for our films were obtained earlier in \cite{sidorova2021magnetoconductance}. We use $\tau_\epsilon$ and $p$ as the only field-independent fitting parameters.  The fit was a subject of two additional restrictions. One requires $\tau_\epsilon$, which is associated with the photoresponse time, to be larger than the phonon escape time. The other requires $\tau_\epsilon$ to follow a power law with respect to temperature. The best-fit dependences $\tau_\epsilon(T_{e0})$ obtained under these conditions with the least-squares method are shown in fig.~\ref{fig:taus}. 
Best-fit values of exponents $p=4.2$ and 3.6 were found for wires with thicknesses $d=6$~nm and 9~nm, respectively. Fitting $\tau_\epsilon(T_{e0})$ data with a power law  (solid curves in fig.~\ref{fig:taus}) results in dependences $\tau_\epsilon \propto T^{-1.9}$ for the 6~nm-thick film and $\tau_\epsilon \propto T^{-1.4}$ for the 9~nm-thick film. We further use the result of the non-linear 2-T photoresponse model and the parameters of our NbTiN films, i.e. $\tau_{EP}$ (weighted relaxation time of the electron energy via phonons), $\tau_{esc}$ and $c_e$ \cite{sidorova2021magnetoconductance}, along with the best-fit values of $\tau_\epsilon(T_{e0})$ obtained here to compute phonon heat capacities at phonon temperatures $T_{ph0}$. The derived $c_{ph}(T_{ph0})$ are shown in fig.~\ref{fig:cphs}.

\section{Discussion}
The studied NbTiN films have polycrystalline granular morphology \cite{Zichi:19} representing crystalline grains embedded in an amorphous environment. Phonon modes are confined in these grains, their wavevectors are restricted in all three directions by grain sizes (following the literature we refer to such structure as having zero dimensionality,  0D). To describe the experimental $c_{ph}$, we employ the simplest model of a cubic grain with the clamped surface (see Appendix~B). Contrary to a free surface, where surface phonon modes exist and enhance the volumetric heat capacity compared to bulk 3D material, surface phonon modes can not emerge on a clamped surface. In fig.~\ref{fig:cphs}, solid curves display fits according to the model with clamped grain surfaces (eq.~(\ref{eq:c_particle})), which we obtained using the number of primitive cells in each direction, $q_a$, as the only fitting parameter (best-fit values $q_a = 4$ and 7). We used the values of sound velocity $u_s = 2.02$~nm/ps and 2.13~nm/ps for NbTiN films with $d=6$~nm and 9~nm, respectively, which were obtained in \cite{sidorova2021magnetoconductance}. As discussed there, these velocities are as twice as less than sound velocities known for crystalline bulk material (calculated from first principles in e.g. \cite{arockiasamy2016ductility}). Reduced velocities are most likely the result of phonon softening caused by the weakening of ion bonds at grain boundaries, defects, and film surfaces \cite{lang2005finite, ma2010electrical}. For comparison, we plot the Debye predictions for a bulk 3D material $c_{ph}^{<3D>}=(2/5)\pi^2k_B^4T^3/(\hbar u_s)^3$ (dashed curves in fig.~\ref{fig:cphs}) using the
sound velocities ($u_s = 2.02$~nm/ps and 2.13~nm/ps for the wires with $d=6$~nm and 9~nm) from  \cite{sidorova2021magnetoconductance}. 
It is clearly seen that both the magnitude and the temperature dependence of experimental $c_{ph}(T)$ differ from the Debye prediction. Specifically, the phonon heat capacity is reduced and the reduction is stronger in the thinner film. We attribute such behavior to the size effects in the phonon heat capacity, which occurs in nanostructures with reduced dimensionality \cite{mcnamara2010quantum}, i.e., when $c_{ph}$ strongly depends on the total  number of allowed vibrational phonon modes.

For both films at large temperatures, the experimental $c_{ph}$ data points deviate from the model predictions (solid lines) for a clamped grain. Although the exact reason of this deviation is not clear at the time, we speculate that the surface clamping might be not complete and depend on temperature.
With regard to the property of the grain surface, it is more plausible that their surfaces are clamped due to the surrounding amorphous environment. According to the model fit, the mean grain size in our NbTiN films is about 1.7~nm and 3~nm for the film with $d=6$~nm and 9~nm, respectively. It was determined as $L=q_a a_0$, where $a_0=0.43$~nm is the NbTiN lattice constant \cite{arockiasamy2016ductility}.

\section{Conclusion}

In this study, the model of the self-heating normal domain was employed for the first time to describe the experimental hysteresis currents in a magnetic field. The impact of the magnetic field on different material parameters and on the shape of the normal domain was analyzed. Such an approach allowed us to retrieve temperature-dependent phonon heat capacities in disordered NbTiN films with polycrystalline granular morphology.

The major results indicate that the phonon heat capacities at low temperatures ($<10$~K) decrease faster with temperature than it is predicted by the Debye model ($c_{ph} \propto T^3$) while its magnitude is well below the Debye predictions and drops with the films' thickness. These findings can be attributed to size effects on the phonon heat capacity and explained by confinement of phonon modes in cubic crystalline grains with clamped surfaces.

Despite the fact that the utilized approach is rather rudimentary and does not capture the full complexity of the NbTiN films under study, it provides valuable insights into the thermal transport in granular films at low temperatures, i.e., (i) confinement of phonon modes in grains, (ii) absence of surface modes due to the clamped surface of grains, and (iii) the effective mean size of grains smaller than the film thickness. Understanding and controlling the size effect as well as boundary properties (i.e., clamped/suspended with absent/present surface modes) are crucial for the design and optimization of micro/nanoscale devices and materials for various applications in superconducting electronics and circuits.

\section{Acknowledgment}
The authors greatly acknowledge V. Zwiller for his help in the sample preparation and A. Sergeev for his essential insights regarding the impact of the magnetic field on electron-phonon interaction.
M.S. acknowledges funding support from the National Research Foundation, Singapore and A*STAR under the Quantum Engineering Programme (QEP-P1).

\appendix 
\section{Self-heating normal domain model}
\label{app_SHNM}

The electron temperature distribution $T_e(x)$ along the wire with a steady-state normal domain is given by the one-dimensional heat-balance equation for unit volumes of the normal and superconducting parts of the wire
\begin{equation}
	- \frac{\partial}{\partial x}\left(\lambda\frac{\partial T_e}{\partial x}\right) + K \left(T_e^p - T_B^p\right) = \frac{I_r^2 R_\square}{w^2d}, \\ 
	\label{eq:norm_dom}
\end{equation}
where the right-hand side of the equation equals zero for the superconducting part. For the present analysis, the electron thermal conductivity $\lambda$ as well as the effective thermal conductance $K=c_e(T_e)/[p \tau_\epsilon(T_e)T_e^{p-1}]$ are taken at $T_e$. Both are assumed to be the same in the superconducting and in the normal state and to be temperature independent. After partial integration and application of boundary conditions (see the details in \cite{sidorova2022phonon}), eq.~\ref{eq:norm_dom} can be reduced to the following algebraic equation valid at the domain center.
\begin{eqnarray}\label{eq:t_J_t_B_pr} 
	T_B^{p+1} + \dfrac{p+1}{p} T_J^p T_C(B) - (T_J^p + T_B^p)^{(p+1)/p} = 0,
\end{eqnarray}
where $T_C(B)$ is the transition temperature in the applied magnetic field and $T_J^p = I_r^2 R_\square / (K w^2 d)$. From eq.~\ref{eq:norm_dom} with the additional boundary condition  $\partial^2 T_e/\partial x^2=0$ at the center of the normal domain, which follows from the translational symmetry, the electron temperature there can be expressed  as $T_{e0}^p = T_J^p + T_B^p$. Substitutions of this expression into the definition of 
$T_J^p$ above and into the eq.~\ref{eq:t_J_t_B_pr} lead to eqs.~(\ref{eq:current}) and (\ref{eq:t_J_t_B}), respectively, in the main text.  
For large domains due to translational symmetry $T_{e0}$ does not depend on the domain size.

\section{Phonon heat capacity}

In general, the phonon heat capacity per unit volume can be expressed as \cite{ehrenreich1977solid}
\begin{equation}\label{eq:c_general}
	c_{ph} = \dfrac{1}{V}  \dfrac{\partial}{\partial T} \sum_{\textbf{k}}  \hbar \omega(\textbf{k}) f_{\textbf{k}},
\end{equation}
where $\textbf{k}$ is the phonon wave vector with components $k_x$, $k_y$, and  $k_z$  in the $x-$, $y-$ and $z$-directions, $\hbar$, $\omega$, and $f_{\textbf{k}}$ is the reduced Planck constants, phonon angular frequency, and Bose-Einstein distribution, respectively. The partial derivative is the temperature derivative of energy of  $f_{\textbf{k}}$ phonons with allowed wavevectors $\textbf{k}$. For bulk samples the set of discrete wave vectors summed over in Eq.~\ref{eq:c_general} becomes dense and the summation can be replaced by an integral. With the linear dispersion, $\omega=u_s k$, this leads to the Debye result $c_{ph}^{<3D>}=(2/5)\pi^2k_B^4T^3/(\hbar u_s)^3$, where $u_s$ is the average sound velocity.

When the dimensionality of the system is reduced as it is in thin films (2D), nanowires (1D), or grains (0D), the replacement of summation in eq.~\ref{eq:c_general} by integration is not valid and the general expression should be used. At low temperatures, surface (planar) modes, for which one of the wavevector components is zero, dominantly contribute to the total heat capacity. When surface modes are allowed, the magnitude of the volumetric phonon heat capacity is enhanced or otherwise suppressed when they are prohibited. Mathematically these two cases are described with different boundary conditions (bc), i.e., the former corresponds to a grain with the free surface (Neumann bc) and the latter to a grain with the clamped surface (Dirichlet bc) \cite{mcnamara2010quantum}.  For a cubic grain with the clamped surface, i.e., without surface modes, the volumetric heat capacity (after rearrangement) is given by \cite{mcnamara2010quantum}
\begin{equation}\label{eq:c_particle}
	c_{ph}^{<0D>} = \dfrac{3 k_B}{a_0^3 q_a^3} \sum_j \sum_m \sum_n \dfrac{\eta^2 e^\eta}{(e^\eta - 1)^2},
\end{equation}
where $\eta = (\pi \hbar u_s / q_a a_0 k_B T) \sqrt{ j^2 + m^2 + n^2}$,  $q_a$ is the number of primitive cells of size $a_0$ in one direction, the indices $j$, $m$, and $n= 1, 2, 3, ..., q_a$. The expression \ref{eq:c_particle} was used to fit experimental data.

\bibliography{manuscript_return_currents/phonon_heat}

\begin{thebibliography}{45}
\bibitem [{com()}]{comment_2TM}%
  \BibitemOpen
  \bibinfo {note} {The uniform two-temperature non-linear model includes two
  time-dependent differential equations for effective electron and phonon
  temperatures valid for large differences between mean, steady state
  temperature values while the photoresponse remains linear, i.e., only small
  deviations from mean values are considered; mathematics of the model will be
  published elsewhere}\BibitemShut {NoStop}%
\end{thebibliography}

\end{document}